\documentstyle [aps,preprint]{revtex}
\tightenlines
\begin{document}
\def\fnote#1#2{
\begingroup\def\thefootnote{#1}\footnote{#2}\addtocounter{footnote}{-1}
\endgroup}
\def\dslash{\not{\hbox{\kern-2pt $\partial$}}}
\def\eslash{\not{\hbox{\kern-2pt $\epsilon$}}}
\def\Dslash{\not{\hbox{\kern-4pt $D$}}}
\def\Aslash{\not{\hbox{\kern-4pt $A$}}}
\def\Qslash{\not{\hbox{\kern-4pt $Q$}}}
\def\Wslash{\not{\hbox{\kern-4pt $W$}}}
\def\pslash{\not{\hbox{\kern-2.3pt $p$}}}
\def\kslash{\not{\hbox{\kern-2.3pt $k$}}}
\def\qslash{\not{\hbox{\kern-2.3pt $q$}}}
\def\np#1{{\sl Nucl.~Phys.~\bf B#1}}
\def\pl#1{{\sl Phys.~Lett.~\bf B#1}}
\def\pr#1{{\sl Phys.~Rev.~\bf D#1}}
\def\prl#1{{\sl Phys.~Rev.~Lett.~\bf #1}}
\def\cpc#1{{\sl Comp.~Phys.~Comm.~\bf #1}}
\def\cmp#1{{\sl Commun.~Math.~Phys.~\bf #1}}
\def\anp#1{{\sl Ann.~Phys.~(NY) \bf #1}}
\def\etal{{\em et al.}}
\def\half{{\textstyle{1\over2}}}
\def\be{\begin{equation}}
\def\ee{\end{equation}}
\def\ba{\begin{array}}
\def\ea{\end{array}}
\def\tr{{\rm tr}}
\title{Newtonian {\em vs} black-hole scattering
\thanks{Research
supported by the DoE under grant DE--FG05--91ER40627.}}
\author{
George Siopsis
\fnote{\dagger}{\tt gsiopsis@utk.edu}}
\address{Department of Physics and Astronomy, \\
The University of Tennessee, Knoxville, TN 37996--1200.\\
}
\date{May 1998}
\preprint{UTHEP--98--0501S}
\maketitle
\begin{abstract}
We discuss non-relativistic scattering by a Newtonian potential.
We show that the gray-body factors associated with scattering by a
black hole exhibit the same functional dependence as scattering amplitudes
in the Newtonian limit, which should be the
weak-field limit of any quantum theory of gravity.
This behavior arises independently of the presence of supersymmetry.
The connection to two-dimensional conformal field theory is also discussed.
\end{abstract}
\renewcommand\thepage{}\newpage\pagenumbering{arabic}

\section{Introduction}

Although the thermodynamic properties of black holes have been
understood for some time~\cite{ref1,ref2,ref3,ref4}, their microscopic origin was only recently
illuminated. This was achieved with the aid of (super)string theory, where
one should be able to count the fundamental degrees of freedom and arrive
at an expression for the entropy~\cite{ref5,ref6,ref7,ref8,ref9}. The semi-classical result
(Bekenstein-Hawking entropy) was thus confirmed. The degrees of freedom
turned out to be solitonic states (D-branes~\cite{ref10}) and not fundamental strings at
all. One could then construct an effective theory by expanding around these
solitons~\cite{ref5,ref11}. The result was a (super)conformal field theory whose validity
extended in the domain of M-theory.

It was later discovered that this effective conformal field theory was more
robust than originally expected. Indeed, not only did it give an accurate
prediction for the entropy, but also for the so-called gray-body factors~\cite{ref12,ref13,ref13a,ref14,ref15}.
This was rather surprising, because there is no apparent connection
between the semi-classical derivation of these factors and the corresponding
analysis in conformal field theory. The two theories (Einstein-Maxwell
gravity and superstring theory) appear to share common properties, pointing
to the existence of a yet-to-be-discovered underlying principle on which
to build a quantum theory of gravity (possibly unified with all the other
forces). In our quest for such a theory, it is important that we derive
expressions for physical quantities (entropy, scattering amplitudes, etc)
under as broad assumptions as possible. What is of interest is the
qualitative behavior of physical quantities, since the fundamental theory
is not known yet.

In this spirit, we consider
the weak-field limit of Einstein gravity, which is, of course, Newtonian
mechanics. We show that the scattering amplitudes in this non-relativistic
limit exhibit the same behavior as the gray-body factors one obtains in
the black-hole background. We conclude that this behavior is more generic
than black-hole backgrounds that can be obtained from D-branes. This sheds
some light on the origin of the gray-body factors, but offers no explanation
on their similarity to factors obtained from (super)conformal field
theory. It would be interesting to see if there is a more direct connection
between non-relativistic scattering and conformal field theory.

Our discussion is organized as follows. In Section~\ref{sec2}, we
derive the gray-body factors in a black-hole background. We perform the
calculation in four dimensions (Schwarzchild metric) and five dimensions
(Kaluza-Klein black holes). In Section~\ref{sec3}, we calculate
scattering amplitudes for a Newtonian potential and show the similarities
of the results with the gray-body factors. Finally, in Section~\ref{sec4},
we discuss our results and their connection to conformal field theory.

\section{Black holes}
\label{sec2}

In this Section, we calculate the gray-body factors associated with Schwarzchild
and Kaluza-Klein black holes. 
We shall do the calculation in the limit where the wavelength of the particle
is much larger than the size of the black hole (small frequency limit).
Similar calculations have already been performed by Maldacena and
Strominger~\cite{ref13}. They solved the wave equation in a black-hole
background by finding solutions in two asymptotic regimes (far from and
near the horizon, respectively) and then matching the solutions. Our solution
is a variant of their method.

\subsection{Schwarzchild black holes}

%
The metric for a Schwarzchild black hole is (we set $c=\hbar = 1$, but
keep $G$)
\begin{equation}
  \label{eq6}
ds^2 = - \left( 1 - {r_0\over r} \right)
\; dt^2 + \left( 1 - {r_0\over r} \right)^{-1}\; dr^2 +
r^2 d\theta^2
+ r^2 \; \sin^2\theta \; d\phi^2
\end{equation}
where $r_0 = 2GM$ is the radius of the horizon.
We wish to study the wave equation for a massless scalar $\Psi$,
\begin{equation}
  \label{eq7}
\Box \Psi = 0
\end{equation}
Using separation of variables, we write $\Psi = R(r) \Theta (\theta) \Phi (\phi)
e^{-i\omega t}$.
The angular part
is the same as in the case of the non-relativistic Schr\"odinger Equation
with a central potential. The radial equation is
\begin{equation}
  \label{eq8}
- {1\over r^2} \; \left(1-{r_0\over r}\right) {d\over dr} \left( r(r-r_0)\; {dR_\ell\over dr} \right) + {\ell(\ell+1)\over r^2}
\; \left(1-{r_0\over r}\right)\;
R_\ell =
k^2 \; R_\ell
\end{equation}
where $k = \omega$ is the wavenumber of the massless scalars.
A massive scalar is described by the same equation, but in that case,
$k = \sqrt {\omega^2 - m^2}$.

To solve this equation, first we need to study the behavior of the
solution in the two asymptotic limits away from the
horizon ($r\to\infty$) and near the horizon ($r\to r_0$),
respectively. More specifically, the two regions will be defined as
$r >> r_0$ and $r << 1/k$, respectively. Notice that the two regions
overlap in the range $r_0 << r << 1/k$, which is non-vanishing
in the small frequency limit $kr_0 << 1$.
We shall solve the wave equation in these two limits and then
match the solutions.

Away from the horizon ($r >> r_0$), we can drop all terms $o(r/r_0)$. Then we can
write Eq.~(\ref{eq8}) as
\begin{equation}
  \label{eq8a}
- {1\over r^2} \; {d\over dr} \left( r^2\; {dR_\ell\over dr} \right) + {\ell(\ell+1)\over r^2}
\; 
R_\ell =
k^2 \; R_\ell
\end{equation}
whose solution is given in terms of a Spherical Bessel function of the
first kind,
\begin{equation}
  \label{eq8b}
R_\ell = A_\ell \; j_\ell (kr)
\end{equation}
where we discarded the solution which is singular in the limit $r\to 0$.
At infinity, this behaves as $R_\ell \sim A_\ell\;
{\sin (kr-\ell \pi/2)\over kr}$. The incoming wave
is $\Psi_{in} = {i^{\ell +1} A_\ell\over 2kr}\;
e^{-ikr}$, so the incoming flux is
\begin{equation}
  \label{eq8c}
J_{in} = -2\pi i r^2\; \left( \Psi_{in}^\ast \; {d\Psi_{in}\over dr} - c.c. \right)
= {|A_\ell|^2 \over k}
\end{equation}
Normalizing to unity, we obtain $A_\ell = \sqrt k$. Therefore,
Eq.~(\ref{eq8b}) becomes
\begin{equation}
  \label{eq8bb}
R_\ell =  \sqrt k\; j_\ell (kr)
\end{equation}
To solve the wave equation near the horizon ($r << 1/k$), define

\begin{equation}
  \label{eq10b}
R_\ell (r) = \left( 1 - {r_0\over r} \right)^{ikr_0} j_\ell (kr) f_\ell (r)
\end{equation}
After some algebra, we obtain for $f_\ell$,
%
\begin{equation}
  \label{eq11}
\left( 1-{r_0\over r} \right) {r\over r_0} {d\over dr} \left( r^2 {df_\ell\over dr} \right)
+ {\cal A} (r) \; {r^2\over r_0} \; {df_\ell\over dr} + {\cal B} (r) \; f_\ell = 0
\end{equation}
where
$${\cal A} (r) = (1+2ikr_0) {r_0\over r} + 2\; \left( 1-{r_0\over r}
\right)
\; {r\over j_\ell(kr)}\; {dj_\ell (kr)\over dr}$$
\begin{equation}
  \label{eq11a}
{\cal B} (r) = - \ell (\ell+1) +(kr_0)^2 \left( 2{r^2\over r_0^2} + 4
{r\over r_0} - 1\right) + (1+2ikr_0) {r\over j_\ell(kr)}\; {dj_\ell
(kr)\over dr}
\end{equation}
To derive the form of the coefficient ${\cal B}$ we made use of
Eq.~(\ref{eq8a}) satified by the Bessel function $j_\ell (kr)$.
Eq.~(\ref{eq11}) is merely a re-writing of the wave equation and no approximations were
performed in deriving it. To proceed further, we need to obtain the form of
the coefficients ${\cal A}$ and
${\cal B}$ in the limit $kr << 1$.
To do this, first expand the Bessel function,
\begin{equation}
  \label{eq11c}
j_\ell (kr) = {(kr)^\ell\over (2\ell +1)!!} \; \Big( 1+ o((kr)^2) \Big)
\;,\quad {dj_\ell (kr) \over dr} = {\ell k^\ell r^{\ell -1}\over (2\ell
+1)!!} \; \Big( 1+ o((kr)^2) \Big)
\end{equation}
We can also discard the terms of the form
$(kr_0)^2(r/r_0)^n << (kr_0)^{2-n}$ ($n=0,1,2$), since $kr_0 < < 1$.
Thus, the coefficients may be written in this limit as
$${\cal A} (r) = 2\ell + (-2\ell+1+2ikr_0) {r_0\over r}$$
\begin{equation}
  \label{eq11aa}
{\cal B} (r) =  \ell (-\ell + 2ikr_0)
\end{equation}
Therefore, near the horizon ($r << 1/k$), Equation~(\ref{eq11}) becomes
\begin{equation}
  \label{eq11ea}
\left( 1-{r_0\over r} \right) {r\over r_0} {d\over dr} \left( r^2 {df_\ell\over dr} \right)
+ \left( 2\ell + (2\ell+1+2ikr_0) {r_0\over r}\right) {r^2\over r_0} \; {df_\ell\over dr} + \ell (-\ell + 2ikr_0) \; f_\ell = 0
\end{equation}
To solve Eq.~(\ref{eq11ea}), we switch variables to $z=r_0/r$. We
obtain
\begin{equation}
  \label{eq11e}
z (1-z) \; {d^2f_\ell\over dz^2} - (2\ell +(-2\ell+1+2ikr_0)\; z ) \; {df_\ell
\over dz}
-\ell (\ell -2ikr_0)\; f_\ell = 0
\end{equation}
whose solution is given in terms of a hypergeometric function,
\begin{equation}
  \label{eq12}
f_\ell (z) = C_\ell \; {}_2F_1(-\ell+2ikr_0\; , \; -\ell\; ; \;
-2\ell\; ; \; z)
\end{equation}
where
\begin{equation}
  \label{eq12a}
{}_2F_1(a\; ,\; b\; ;\; c\; ;\; z) = 1 +{ab\over c}\; z + {a(a+1)
b(b+1)\over c(c+1)}\; {z^2\over 2!} + {a(a+1)(a+2)b(b+1)(b+2)
\over c(c+1)(c+2)}
\; {z^3\over 3!} + \cdots
\end{equation}
In our case, the series terminates, and $f_\ell$ is a polynomial of degree
$\ell$. Therefore, $f_\ell$ is regular in the entire domain
$0\le z\le 1$, as expected, because we already isolated the
singular behavior of the wavefunction $R_\ell$ in the definition~(\ref{eq10b}).
Transforming back to the radial coordinate $r$,
from Eqs.~(\ref{eq10b}) and (\ref{eq12}), we obtain
\begin{equation}
  \label{eq12c}
R_\ell (r) = C_\ell \; \left( 1- {r_0\over r} \right)^{ikr_0} j_\ell (kr)
{}_2F_1(-\ell+2ikr_0\; , \; -\ell\; ; \; -2\ell\; ; \; r_0/r\, )
\end{equation}
Having obtained the asymptotic form of the solution in the regions $r >> r_0$ (Eq.~(\ref{eq8bb}))
and $r << 1/k$ (Eq.~(\ref{eq12c})), we shall now match the two expressions in the limit $r\to\infty$.
This implies
\begin{equation}
  \label{eq12cc}
C_\ell = \sqrt k
\end{equation}
Next, we calculate the flux at the horizon. 
Near the horizon, $r\to r_0$, so $z\to 1$. The value of the
hypergeometric function at $z=1$ can be obtained from the hypergeometric
identity
\begin{equation}
  \label{eq12b}
{}_2F_1(-\ell+2ikr_0\; , \; -\ell\; ; \; -2\ell\; ; \; z)
= {\textstyle{\Gamma (\ell + 1)\Gamma (\ell +1+2ikr_0)
\over\Gamma (2\ell + 1)\Gamma (1+2ikr_0)}}\;
{}_2F_1(-\ell+2ikr_0\; , \; -\ell\; ; \;1+2ikr_0\; ; \; 1-z)
\end{equation}
Switching to the variable $\xi = 1-z$, near the horizon we obtain
from Eq.~(\ref{eq12c})
\begin{equation}
  \label{eq12d}
R_\ell (\xi ) = \sqrt k\; {\textstyle{\Gamma (\ell + 1)\Gamma (\ell +1+2ikr_0)
\over\Gamma (2\ell + 1)\Gamma (1+2ikr_0)}}\; j_\ell (kr_0) \xi^{ikr_0}
\end{equation}
so the flux at the horizon is
$$J_h^\ell = -2\pi i \; r( r-r_0 )\; \left( R_\ell^\ast
{dR_\ell\over dr} - c.c.\right) = -2\pi i \xi r_0\; \left( R_\ell^\ast
{dR_\ell\over d\xi} - c.c.\right) $$
\begin{equation}
  \label{eq13}
= 4\pi \left( {\ell !\over (2\ell +1)! (2\ell +1)!!} \right)^2
\left| {\Gamma (\ell +1 +2ikr_0) \over \Gamma (1 +2ikr_0)} \right|^2
\; (kr_0)^{2\ell +1}
\end{equation}
where in the last step we made use of Eq.~(\ref{eq11c}).
The gray-body factors (decay rates at the horizon) are given by
\begin{equation}
  \label{eq15}
\Gamma_\ell = {\pi J_h^\ell\over k^2 (e^{4\pi kr_0} - 1)}
= {\pi (\Gamma (\ell +1))^2\over 2^{2\ell +2}
(\Gamma (\ell + {3\over 2}))^2 (\Gamma (2\ell +1))^2}\;
k^{2\ell -1} r_0^{2\ell +1} e^{-2\pi kr_0} |\Gamma (\ell + 1+2ikr_0)|^2
\end{equation}
They may also be written in terms of the Hawking temperature, $T_H = {1\over 4\pi r_0}$ and
horizon area $A = 4\pi r_0^2$,
\begin{equation}
  \label{eq15a}
\Gamma_\ell =
{\pi (\Gamma (\ell +1))^2\over 2^{2\ell +2}
(\Gamma (\ell + {3\over 2}))^2 (\Gamma (2\ell +1))^2}\;
k^{2\ell -1} (T_HA)^{2\ell +1} e^{-k/(2T_H)} |\Gamma (\ell + 1+ik/(2\pi T_H)|^2
\end{equation}

\subsection{Kaluza-Klein black holes}

Consider five-dimensional space-time with an internal periodic fifth dimension $x_5$. The metric in
five dimensions can be split into a four-dimensional metric $g_{\mu\nu}$, gauge field $A_\mu$ and
scalar $\chi$, with dynamics governed by the action~\cite{ref16}
\begin{equation}
  \label{eq16a}
S = {1\over 16\pi G} \int d^4x \sqrt{-g} \Big(R - 2\partial_\mu\chi\partial^\mu\chi
- e^{-2\sqrt 3\chi} F_{\mu\nu} F^{\mu\nu} \Big)
\end{equation}
The momentum along $x_5$ gives rise to a charge. Thus, we obtain charged black hole
solutions with four-dimensional metric,
\begin{equation}
  \label{eq16}
ds^2 = - {1\over \sqrt\Delta} \left( 1-{r_0\over r} \right) dt^2 + \sqrt\Delta
\left( {dr^2\over (1-r_0/r)} + r^2d\theta^2 + r^2\sin^2\theta d\phi^2 \right)
\end{equation}
where $\Delta = e^{-4\chi /\sqrt 3} = 1+{r_0\sinh^2\gamma\over r}$, and gauge field
$A_0 = - {r_0\sinh (2\gamma)\over 4r\Delta}$.
The ADM mass, charge, entropy and Hawking temperature of the black hole, respectively, are
\begin{equation}
  \label{eq17}
M = {r_0\over 8G}\; (3+\cosh (2\gamma)) \;,\quad Q = {r_0\over 4G}\; \sinh (2\gamma)\;,
\quad S = {\pi r_0^2\over G}\; \cosh\gamma \;, \quad T_H = {1\over 4\pi r_0\cosh\gamma}
\end{equation}
The scattering of a neutral massless scalar is described by the wave equation
\begin{equation}
  \label{eq18}
\Delta\; {\partial^2\Psi\over \partial t^2} - {1\over r^2} \; \left(1-{r_0\over r}\right)
{d\over dr} \left( r(r-r_0)\; {d\Psi\over dr} \right) + {1\over r^2}
\; \left(1-{r_0\over r}\right)\; L^2\;
\Psi =0
\end{equation}
The eigenvalues of $L^2$ are $\ell (\ell+1)$, so the radial part of the wavefunction for scalars of
energy $\omega = k$ satisfies
\begin{equation}
  \label{eq18a}
- {1\over r^2} \; \left(1-{r_0\over r}\right) {d\over dr} \left( r(r-r_0)\; {dR_\ell\over dr} \right) + {\ell(\ell+1)\over r^2}
\; \left(1-{r_0\over r}\right)\;
R_\ell =
k^2 \; \Delta\; R_\ell
\end{equation}
Working as before, first we derive the behavior of the
wavefunction in the two asymptotic limits away from the
horizon ($r >> r_0$) and near the horizon ($r << 1/k$),
respectively.
Away from the horizon, working as before, we obtain
\begin{equation}
  \label{eq11eab}
\left( 1-{r_0\over r} \right) {r\over r_0} {d\over dr} \left( r^2 {df_\ell\over dr} \right)
+ \left( 2\ell + (2\ell+1+2ikr_0\cosh\gamma) {r_0\over r}\right) {r^2\over r_0} \; {df_\ell\over dr} + \ell (-\ell + 2ikr_0\cosh\gamma) \; f_\ell = 0
\end{equation}
This only differs from Eq.~(\ref{eq11ea}) by the substitution $kr_0 \to kr_0 \cosh\gamma$.
Notice that, written in terms of the Hawking temperature, both Eqs.~(\ref{eq11ea}) and (\ref{eq11eab}) read
\begin{equation}
  \label{eq19a}
\left( 1-{r_0\over r} \right) {r\over r_0} {d\over dr} \left( r^2 {df_\ell\over dr} \right)
+ \left( 2\ell + \left(2\ell+1+i\; {k\over 2\pi T_H}\right) {r_0\over r}\right) {r^2\over r_0} \; {df_\ell\over dr} + \ell \left(-\ell + i\; {k\over 2\pi T_H}\right) \; f_\ell = 0
\end{equation}
The rest of the calculation proceeds along the same lines as the derivation
of the Schwarzchild gray-body factors in the small frequency limit~(\ref{eq15}).
In this case, we obtain
\begin{equation}
  \label{eq20}
\Gamma_\ell = 
{\pi (\Gamma (\ell +1))^2\over 2^{2\ell +2}
(\Gamma (\ell + {3\over 2}))^2 (\Gamma (2\ell +1) )^2}\;
k^{2\ell -1} r_0^{2\ell +1} (\cosh\gamma)^{2\ell +1}
e^{-2\pi kr_0\cosh\gamma} |\Gamma (\ell + 1+2ikr_0\cosh\gamma)|^2
\end{equation}
It terms of the Hawking temperature (Eq.~(\ref{eq17})),
we obtain
\begin{equation}
  \label{eq20a}
\Gamma_\ell \sim
k^{2\ell -1} e^{-k/(2T_H)} |\Gamma (\ell + 1+ik/(2\pi T_H)|^2
\end{equation}
which is of the same form as Eq.~(\ref{eq15a}).

This can be generalized to a four-dimensional black hole obtained from string theory. To do this, we start
with ten-dimensional spacetime and compactify the six dimensions on a torus~\cite{ref5}. The four-dimensional metric is
given by Eq.~(\ref{eq16}), where
\begin{equation}
  \label{eq21}
\Delta = f(\gamma_1) f(\gamma_2) f(\gamma_3) f(\gamma_4) \;, \quad f(\gamma_i) = 1+{r_0
\sinh^2\gamma_i\over r}
\end{equation}
The parameters $\gamma_i$ ($i = 1,\dots ,4$)
are related to the charges of the black hole.
The ADM mass, entropy and Hawking temperature of the black hole, respectively, are ({\em cf.}~
Eq.~(\ref{eq17}))
\begin{equation}
  \label{eq22}
M = {r_0\over 8G}\; \sum_{i=1}^4 \cosh (2\gamma_i) \;,
\quad S = {\pi r_0^2\over G}\; \prod_{i=1}^4\cosh\gamma_i \;, \quad T_H = {1\over 4\pi r_0}\; \prod_{i=1}^4
{1\over\cosh\gamma_i} 
\end{equation}
Working as before, we arrive at the same results in the small frequency
limit, provided we substitute
$r_0 \to r_0\cosh\gamma_1\cosh\gamma_2\cosh\gamma_3\cosh\gamma_4$.
Once again, the gray-body factors are of the same form~(\ref{eq20a}), where $T_H$ is given by Eq.~(\ref{eq22}).

Next, we consider the non-relativistic limit of Newtonian scattering. Even though there is no horizon present, the scattering amplitudes exhibit the same
behavior near the center of gravity.


\section{Newtonian scattering}
\label{sec3}

Consider a heavy body of mass $M$ ({\em e.g.}~the black hole of the
previous Section) and an incident beam of light particles of
reduced mass $m$. The particles have (non-relativistic)
relative speed $v = k/m$ in the $z$-direction.
Their scattering by the heavy body is described by the
non-relativistic Schr\"odinger Equation
\begin{equation}
  \label{eq1}
-\; {1\over 2m} \; \nabla^2 \Psi +{GMm\over r} \; \Psi = E \Psi
\end{equation}
where $E = k^2/2m$. This Equation can be solved exactly for the given boundary conditions,
by using parabolic coordinates. Normalizing the incident flux to unity, we obtain
\begin{equation}
  \label{eq1a}
\Psi = {1\over \sqrt v} \; \Gamma (1+i\eta) e^{-\pi\eta/2} e^{ikz} F(-i\eta\;,\; 1\; ;\; 2ikr\sin^2
{\textstyle{1\over 2}} \theta)
\end{equation}
where $\eta = GMm^2/k$ and $F$ is the hypergeometric function
\begin{equation}
  \label{eq1b}
F(a,b;z) = 1 +{a\over b}\; z + {a(a+1)\over b(b+1)}\; {z^2\over 2!} + {a(a+1)(a+2)\over b(b+1)(b+2)}
\; {z^3\over 3!} + \cdots
\end{equation}
This solution can be expanded in partial waves,
\begin{equation}
  \label{eq2}
\Psi = \sum_{\ell =0}^\infty R_\ell (r) P_\ell (\cos\theta)
\end{equation}
where the $R_\ell$ satisfy the Radial Equation
\begin{equation}
  \label{eq3}
-\; {1\over r^2} {d\over dr} \left( r^2 {dR_\ell\over dr} \right) +\left(
{2\eta k\over r} + {\ell(\ell+1)\over r^2} \right)\; R_\ell = k^2R_\ell
\end{equation}
We obtain
\begin{equation}
  \label{eq3a}
R_\ell (r) = {1\over \sqrt v} \; {\Gamma (\ell + 1+i\eta)\over \Gamma (2\ell +1)}\;
e^{-\pi\eta/2} \; (2ikr)^\ell \; e^{ikr} F(\ell + 1+i\eta\;,\; 2\ell +2\; ;\; -2ikr)
\end{equation}
Close to the center of gravity ($r\to 0$), we obtain
\begin{equation}
  \label{eq4}
|R_\ell|^2 \sim {1\over v} \; {|\Gamma (\ell + 1+i\eta)|^2\over (\Gamma(2\ell+1))^2}\;
(2kr)^{2\ell} e^{-\pi\eta} = {m\over \hbar} \; {2^{2\ell}\over
(\Gamma(2\ell+1))^2}\; e^{-\pi\eta} k^{2\ell-1} r^{2\ell}
|\Gamma (\ell + 1+i\eta)|^2
\end{equation}
In particular, the particle density at the center of gravity is found from Eq.~(\ref{eq1a}), if we set $r=0$,
\begin{equation}
  \label{eq5}
|\Psi (0)|^2 = {1\over v} \; |\Gamma (1+i\eta)|^2 e^{-\pi\eta} = {1\over v} \; {2\pi\eta\over
e^{2\pi\eta} -1}
\end{equation}
This may be viewed as blackbody spectrum of the wavenumber $k$ at temperature
\begin{equation}
  \label{eq5a}
T = {k\over 2\pi\eta} = {v^2\over 2 \pi GM}
\end{equation}
The ensemble consists of particles of varying masses and wavenumbers, but
of constant incoming speed.
If we express the partial-wave rates (Eq.~(\ref{eq4})) in terms of this
temperature, we obtain
\begin{equation}
  \label{eq4a}
|R_\ell|^2 = m \; {2^{2\ell}\over
(\Gamma(2\ell+1))^2}\; e^{-k/ 2T} k^{2\ell-1} r^{2\ell}
|\Gamma (\ell + 1+i k/ (2\pi T))|^2
\end{equation}
This is of the same functional form as the gray-body factors $\Gamma^\ell$
(Eq.~(\ref{eq15})) that were derived by using the exact black-hole
potential. Of course, the temperature in the non-relativistic case is
arbitrary, because there is no horizon effect. Still, the similarity with
Eq.~(\ref{eq15a}) is non-trivial.
It should also be pointed out that the differential equations in the two
cases are different; their respective solutions are expressed in terms
of different hypergeometric functions.

\section{Discussion}
\label{sec4}

The microscopic calculation of the entropy of black holes in superstring
theory has left little doubt that strings hold the key to the discovery
of the quantum theory of gravity. On the other hand, the fact that the
microscopic calculation involves the counting of solitonic states
($D$-branes) shows that a more fundamental theory is needed that will
provide the missing underlying principle on which quantum gravity should
be based.

The first microscopic calculation~\cite{ref5} seemed to rely heavily on
supersymmetry. The agreement between the microscopic and macroscopic
calculations was guaranteed by supersymmetry, which ensured that the
number of supersymmetric (BPS) states was invariant when the string
coupling was varied. It was therefore surprising to discover that there
was agreement between the two approaches that went beyond the demands of
supersymmetry. Such an agreement was demonstrated at a fairly detailed
level with non-extremal black holes and gray-body factors~\cite{ref12,ref13,ref13a,ref14,ref15}.

We have discussed the behavior of gray-body factors for
non-supersymmetric black holes. The goal was to understand the origin of
their behavior. We have shown that there is a striking agreement with
partial-wave amplitudes in the non-relativistic limit. This agreement
is rather non-trivial as the calculations in the two cases (exact and
non-relativistic approximation) rely on different differential equations
possessing solutions that are expressed in terms of different
hypergeometric functions.
Therefore, the behavior of the gray-body factors seems to be universal.

In the cases we studied, there is no corresponding supestring theory, so
a microscopic calculation is not readily available. However, it should
be pointed  out that one may still derive the functional dependence of
the gray-body factors from conformal field theory. Indeed, if we
introduce the chiral operator $\Theta (\sigma^+)$ of conformal dimension
$\ell +1$, the thermal correlators at temperature $T$ are~\cite{ref13,ref13a}
\begin{equation}
  \label{eq23}
\langle \Theta^\dagger (0) \Theta (\sigma^+) \rangle_T \sim {1\over
\sinh^{2(\ell+1)} \pi T\sigma^+}
\end{equation}
The gray-body factors are
\begin{equation}
  \label{eq24}
\Gamma^\ell \sim \int d\sigma^+ e^{-ik\sigma^+}
\langle \Theta^\dagger (0) \Theta (\sigma^+) \rangle_T
\end{equation}
where we identify $\sigma^+ \equiv \sigma^+ +2i/T$. We obtain
\begin{equation}
  \label{eq25}
\Gamma^\ell \sim e^{-k /2T} |\Gamma (\ell +1+ik/(2\pi T))|^2
\end{equation}
in agreement with our earlier results.

In conclusion, we have shown that there is an agreement between
\begin{itemize}
\item[{\em (a)}] gray-body factors calculated from black-hole dynamics;
\item[{\em (b)}] partial-wave scattering amplitudes in the
non-relativistic limit;
\item[{\em (c)}] thermal correlators in chiral conformal field theory.
\end{itemize}
We found agreement at a fairly detailed level, even though the three
calculations bared little resemblance to one another. It might be
interesting to extend these results to higher space-time dimensions and more general
classes of black holes. Such explorations should shed light on the
yet-to-be-discovered underlying principle of quantum gravity and the
information loss paradox.



\end{document}